\documentstyle[editedvolume,amsthm,amsfonts,amsmath,numreferences]{crckapb} 

\def\req#1{(\ref{#1})}

\newcommand{\Tr}{{\rm Tr}}

\begin{opening}
\title{FULL COUNTING STATISTICS}
\subtitle{An elementary derivation of Levitov's formula}
\author{I. KLICH}
\institute{Technion\\Department of Physics, 32000 Haifa, Israel.}
\end{opening}
\begin{document}
 \maketitle \vskip 2mm

The field of quantum noise has been rapidly developing in recent
years, with the growing possibilities in precision measurements
\cite{BlanterReview}, and interest in mesoscopic systems as well
as in technological applications of physical effects at the
micrometer and nanometer scales.

Of particular interest is the study of the statistics of charge
transport between materials coupled through a contact or through a
time dependent scatterer. The full statistics of charge transport
was studied in a series of works by Levitov et al.
\cite{Levitov,Ivanov}. This approach yielded interesting results,
and in particular, they where able to express the full counting
statistics in terms of a determinant of a single particle
operator. Several aspects of Levitov's formula \req{LevitovForm}
where discussed in following papers \cite{Andreev,Nazarov}.

Our aim in this paper is to present a novel derivation of the
original Levitov formula. This is done by proving a trace formula
\req{formula}, which relates certain traces in Fock space to
single particle determinants. Using the present approach we find
in addition several generalizations, such as a corresponding
formula for Bosons.

\section{The full counting statistics}

The typical setting is the following: consider particle
reservoirs, with given temperatures and chemical potentials, which
are separated at time zero, and are evolving by a second quantized
hamiltonian $H_0$. At some time the reservoirs are coupled,
through a scattering region, and evolve by a new, time dependent
hamiltonian. After a time $T$ they are decoupled again and one is
interested in the statistics of charge transported from side to
side, i.e. to compute $<Q>,<Q^2>$ and higher moments. The role of
the third moment was recently discussed in \cite{LevitovReznikov}.

The full statistics of charge transfer may be conveniently
represented by the characteristic function of the (charge
transport) probability distribution function, defined by:
\begin{equation}
\chi(\lambda_1,...,;T)=\sum_{{\bf \alpha},{\bf \beta}}P({\bf
\alpha}(t=0),{\bf
\beta}(t=T))e^{iq\sum_{i}\lambda_i(\beta_i-\alpha_i)}
\end{equation}
Here the summation is over all states ${\bf
\alpha}=(\alpha_1,...),{\bf \beta}=(\beta_1,...)$ labelling the
Fock space in the occupation number representation: for fermions
these are vectors of zeros and ones and for bosons vectors with
integer coefficients, where $\alpha_i$ is the number of particles
occupying the single particle state $i$ in ${\bf\alpha}$. $P({\bf
\alpha}(t=0),{\bf \beta}(t=T))$ is the probability that we started
in state ${\bf \alpha}$ at time $0$ and finished in a state ${\bf
\beta}$ at time $T$. Thus terms $(\alpha_i-\beta_i)$ appearing in
the exponent are just the change in the number of particles
occupying the single particle state $i$. And the parameters
$\lambda_i$ are introduced in the standard manner to calculate
different moments. By taking derivatives of $\chi$ with respect to
$\lambda_i$, one can calculate arbitrary moments of the charge
accumulation in state $i$. For example,
\begin{equation}
<Q_i>=-i\partial_{\lambda_i}\log\chi|_{\lambda_1,...=0}
\end{equation}
And
\begin{equation}
(\Delta
Q_i)^2=<Q_i^2-<Q_i>^2>=-\partial_{\lambda_i}^2\log\chi|_{\lambda_1,...=0}.
\end{equation}

 It is also possible to compute the moments of the charge
which is transferred to a particular reservoir by taking the
derivative with respect to $\lambda$ after setting
$\lambda_i=\lambda$ for all states $i$ belonging to the desired
reservoir. $\chi(\lambda)$ may be interpreted as the reaction of
the system to coupling with a classical field $\lambda$ which
measures the number of electrons on each side.

For adiabatic change, and short scattering time Levitov et al
\cite{Levitov}, obtained the following expression for $\chi$:
\begin{eqnarray}\label{LevitovForm}
\chi({\lambda})=\det(1+n(S^{\dag}e^{iq\lambda}Se^{-iq\lambda}-1))
\end{eqnarray}
Where $n$ is the occupation number operator and $S$ is the
scattering matrix. As was remarked \cite{Levitov}, this expression
requires careful understanding and regularization. In the
following we derive this formula in a new manner which, we hope
will allow a convenient way to address these issues.

In order to proceed we first write $\chi$ as a trace in Fock
space:
\begin{eqnarray}\label{chitr}
& &\chi({\lambda},T) =\\ \nonumber & &\sum_{{\bf \alpha},{\bf
\beta}}<{\bf \alpha}|\rho_0|{\bf \alpha}>|<{\bf
\alpha}|{\mathbb U}^{\dag}|{\bf \beta}>|^2e^{iq\sum_i\lambda_i(\beta_i-\alpha_i)}\\
\nonumber & & ={\rm Tr}(\rho_0 {\mathbb
U}^{\dag}e^{iq\sum\lambda_i a^{\dag}_i a_i}{\mathbb
U}e^{-iq\sum\lambda_i a^{\dag}_i a_i})
\end{eqnarray}
Where $\rho_0$ is the density matrix at the initial time ($t=0$),
 ${\mathbb U}$ is the evolution (in Fock space) from time $t=0$ to
time $t=T$ and $a^{\dag}_i ,a_i$ are the creation and annihilation
operators for a given one particle state $i$. Here it is assumed
that the occupation number basis is chosen such that the initial
time density matrix $\rho_0$ is diagonal in it, which implies that
the states $\bf \alpha$ are eigenstates of the initial
Hamiltonian, and measurement of charge in a specific state is
meaningful.

Next, we define the second quantized version of a single particle
operator $A$ (i.e. an operator on the single particle Hilbert
space) to be the Fock space operator:
\begin{equation}\label{secondquantized}
\Gamma(A)=\sum <i|A|j>a^{\dag}_i a_j.
\end{equation}
Then \req{chitr} can be
written as:
\begin{eqnarray}
\chi({\lambda},T)={\rm Tr}(\rho_0 e^{iq{\mathbb U}^{\dag}\Gamma
({\bf \lambda}){\mathbb U}}e^{-iq\Gamma({\bf \lambda})})
\end{eqnarray}
Here ${\bf \lambda}$ is the matrix ${\rm
diag}(\lambda_1,\lambda_2,...)$. To handle this kind of
expressions (and to obtain Levitov's formula) we prove in the
following section a trace formula.

\section{A trace formula}
In this section we prove the following:
\begin{equation}\label{formula}
{\rm Tr}( e^{\Gamma(A)}e^{\Gamma(B)})={\rm det}(1-\xi
e^{A}e^{B})^{-\xi}
\end{equation}
Where $\xi=1$ for bosons and $\xi=-1$ for fermions (i.e. the
creation and annihilation operators satisfy $a_j a^{\dag}_i-\xi
a^{\dag}_i a_j=\delta_{ij}$).

We prove this result for the finite dimensional Hilbert space
case, and avoid at this point questions regarding the limit of
infinite number of states, to be addressed elsewhere \cite{AK}.
\\
{\it Proof}: \\ For an $N$ dimensional single particle Hilbert
space $\Gamma$ is a representation of the usual Lie algebra of
matrices $gl(N)$. Indeed, substituting the definition
\req{secondquantized}, together with the relations obeyed by the
creation and annihilation operators it is straightforward to check
that
\begin{eqnarray}\label{representation}
& [\Gamma(A),\Gamma(B)]=\Gamma([A,B])
\end{eqnarray}
is true for bosons and for fermions. By Baker Campbell Hausdorf
there exists a matrix $C$ such that $e^A e^B=e^C$. $C$ is an
element of $gl(N)$ and is given by a series of commutators, since
$\Gamma$ is a representation, it holds that
\begin{equation}
e^A e^B=e^C\rightarrow e^{\Gamma(A)}e^{\Gamma(B)}=e^{\Gamma(C)}.
\end{equation}
Now let us evaluate ${\rm Tr}( e^{\Gamma(C)})$. Any matrix $C$ can
be written in a basis in which it is of the form $\rm
diag(\mu_1,..\mu_n)+K$ where $K$ is an upper triangular, thus we
have
\begin{eqnarray}
& {\rm Tr}( e^{\Gamma(C)})= {\rm Tr}( e^{\Gamma({\rm
diag}(\mu_1,..,\mu_n))+\Gamma(K)})={\rm Tr}( e^{\Gamma({\rm
diag}(\mu_1,..,\mu_n))})=\\ \nonumber & {\rm Tr}( \prod_i e^{\mu_i
a^{\dag}_ia_i})=\prod_i (1-\xi e^{\mu_i})^{-\xi}={\rm det}(1-\xi
e^{C})^{-\xi}
\end{eqnarray}
(One may also think of ${\rm Tr}( e^{\Gamma(C)})$ as the partition
function of a system with Hamiltonian $-C$ at temperature $k_B
T=1$). From this equation \req{formula} follows:
\begin{equation}
{\rm Tr}( e^{\Gamma(A)}e^{\Gamma(B)})={\rm Tr}(
e^{\Gamma(C)})={\rm det}(1-\xi e^{C})^{-\xi}={\rm det}(1-\xi
e^{A}e^{B})^{-\xi}
\end{equation}
We remark at this point that this relation can immediately be
generalized in the same way to products of more then two
operators.
\\
$\bullet$ Let us illuminate our identity with a trivial example:
Let ${\cal H}$ be an $N$ - dimensional Hilbert space, and choose
$A=B=0$.
Then the dimension of the appropriate Fock space is given by \\
$\Tr({\mathbb{I}})=\Tr( e^{\Gamma(0)}e^{\Gamma(0)})={\rm
det}(1-\xi)^{-\xi}=\Big\{\begin{array}{lll} & 2^N & {\rm Femions}
\\ \nonumber & \infty & {\rm Bosons}\end{array}$ \\ as it should
be.

\section{Levitov's formula}
We now turn to give a novel derivation of Levitov's result for the
full counting statistics. In the framework of non interacting
fermions the evolution ${\mathbb U}$ in the expression \req{chitr}
for $\chi$ is just the Fock space implementation of the single
particle evolution $U$. That means that ${\mathbb U}^{\dag}\Gamma
(\lambda){\mathbb U}=\Gamma ( U^{\dag}\lambda U)$ so that by the
trace formula \req{formula} for 3 operators, we immediately have
\begin{eqnarray}\label{chieq}
& & \chi({\lambda},T)={\rm Tr}({e^{-\beta \Gamma(H_0)}\over Z}
e^{iq\Gamma ( U^{\dag}\lambda U)}e^{-iq\Gamma({\bf \lambda})})\\
\nonumber & & {1\over
Z}\det(1+e^{-\beta H_0} (U^{\dag}e^{iq\lambda}Ue^{-iq\lambda}))=\\
\nonumber & & \det(1+n (U^{\dag}e^{iq\lambda}Ue^{-iq\lambda}-1))
\end{eqnarray}
Where  $Z=\det(1+e^{-\beta H_0})$ and $n$ is the occupation number
operator ${e^{-\beta H_0}\over 1+e^{-\beta H_0}}$ at the initial
time. We note that the result \req{chieq} should be viewed as the
general expression for the counting statistics of noninteracting
fermions, at any given time, and without any approximation. And
may be a good start for studying different limits of the problem,
as well as regularization difficulties.

Finally, if the scattering time is very small compared with the
entire evolution, then one may describe the problem in terms of
dynamical scattering operators, $S=\lim_{t\rightarrow\infty}
e^{iH_0 t} U(t,-t)e^{iH_0 t}$ where $H_0$ is the initial free
evolution. Using the fact that $\lambda$ commutes with $H_0$, one
obtains in the limit of $T\rightarrow\infty$:
\begin{eqnarray}
\chi({\lambda})=\det(1+n(S^{\dag}e^{iq\lambda}Se^{-iq\lambda}-1))
\end{eqnarray}
Which is Levitov's result \req{LevitovForm}, as promised.
\\  We now add a few remarks:

1. {\it Convergence and regularization:} First we note that as
long as we assume that ${\mathbb \rho}$ exists and has trace 1,
then trace of $\rho$ times a bounded operator is also finite, so
that $\chi$ is well defined. However, problems might arise when
taking the thermodynamic limit. Taking the infinite volume limit
may cause the Fock space density matrix to be ill defined (i.e it
cannot be normalized to trace 1), however, expectation values
obtained using it may still have meaning.

If one uses the scattering matrix as in \req{LevitovForm},
regularization of the determinant is needed, since in this case
one uses the static scattering matrix as an approximation for the
true evolution. Indeed, in the limit $T\rightarrow\infty$
arbitrarily large charges can pass from side to side \footnote{To
see this we note that the first moment of \req{LevitovForm}, which
is the transported charge, diverges as time goes to infinity if
there is a bias between the reservoirs}, so that the information
about the length of the time interval has to be put in by hand
\cite{Levitov}. The equation \req{chieq}, however, can be shown to
be well defined even in the thermodynamic limit, for a finite time
interval \cite{AK}.

2. {\it Bosons:} It is now straightforward to derive an analogous
formula for the full counting statistics of bosons. The result is
simply:
\begin{eqnarray}
\chi_{B}({\lambda},T)={1\over\det(1-n_B(U^{\dag}e^{iq\lambda}Ue^{-iq\lambda}-1))}.
\end{eqnarray}
Where $n_B$ is the occupation number operator for bosons.

3. {\it Rate of charge accumulation:} Here we give an example of
how one may compute the rate of charge accumulation in a box $A$.
We choose a box in space, which can be described by a projection
$P_A$ in the single particle Hilbert space (with matrix elements
$<x|P_A|x'>=\delta(x-x')$ if $x$ is in the box, and zero
otherwise). By setting $\lambda=0$ for all states that are outside
the box $A$, one finds
\begin{eqnarray}
&
&\dot{Q_A}=-i\partial_t\partial_{\lambda}\log\chi({\lambda},t)|_{\lambda=0}=\\
\nonumber & & q\partial_t\Tr(n(U^{\dag}P_A
U-P_A))=q\Tr(n(\dot{U}^{\dag}P_A U+U^{\dag}P_A \dot{U}))=\\
\nonumber & & q\,{\rm Re}<U\dot{U}^{\dag}P_A>_t
\end{eqnarray}
The angular brackets describe averaging over the distribution at
the time of measurement $t$. Charge accumulation is equivalent to
current if the box $A$ is connected via just one contact to the
other reservoirs. This equation should be compared to the formula
for the current in terms of scattering matrices
\cite{BPT,Brouwer,Avron,Makhlin} which is of fundamental interest
in the field of quantum pumps.

\section{summary}
To conclude, we presented a novel derivation of Levitov's
determinant formula for the full counting statistics of charge
transfer. This was done by introducing a trace formula
\req{formula} which is suitable for translating problems of
non-interacting particles from Fock space to the single particle
Hilbert space. The derivation is general enough to allow
consideration of new problems of counting statistics, in
particular, further problems involving bosons, or measurement of
other operators then charge. We hope that some properties of the
determinant \req{chieq} under various limits, such as adiabatic
and thermodynamic limits will now be easier to address.
\\
\\
I am grateful to J. E. Avron for many discussions and remarks, and
especially to O. Kenneth for his help in the proof of
\req{formula}. I would also like to thank J. Feinberg, L. S.
Levitov and M. Reznikov for remarks.
\\
\\

\end{document}